\documentclass[preprint,sort&compress, 5p,times]{elsarticle}

\usepackage{lipsum}
\makeatletter
\def\ps@pprintTitle{%
 \let\@oddhead\@empty
 \let\@evenhead\@empty
 \def\@oddfoot{}%
 \let\@evenfoot\@oddfoot}
\makeatother

\usepackage{float,ulem}
\usepackage{graphicx,epsfig,epstopdf,amsmath,amssymb}
\usepackage{xcolor}
\usepackage[colorlinks=true,linkcolor=blue,citecolor=blue,urlcolor=blue]{hyperref}
\usepackage{caption}
\usepackage{subcaption}

\newcommand{\be}{\begin{equation}}
\newcommand{\ee}{\end{equation}}
\newcommand{\ba}{\begin{eqnarray}}
\newcommand{\ea}{\end{eqnarray}}
\newcommand{\nn}{\nonumber\\}
\def\k{\boldsymbol k}
\begin{document}

\title{Why do newer degrees of freedom appear in higher-order truncated hydrodynamic theory?}

\author{Sukanya Mitra}
\ead{sukanya.mitra@niser.ac.in}
\address{School of Physical Sciences, National Institute of Science Education and Research, An OCC of Homi Bhabha National Institute, Jatni-752050, India.}

\begin{abstract}
An exact derivation of relativistic hydrodynamics from an underlying microscopic theory has been shown to be an all-order theory. From the relativistic transport equation of kinetic theory, the full expressions of hydrodynamic viscous fluxes have been derived which turn out to include all orders of out-of-equilibrium derivative corrections.
It has been shown, that for maintaining causality, it is imperative that the temporal derivatives must include all orders, which can be resummed in non-local, relaxation operator-like forms and finally `integrated in' introducing newer degrees of freedom. The theory can of course be truncated at any higher spatial orders, but the power over the the infinite temporal sum increases correspondingly such that the causality is respected. As a result, the theory truncated at any higher order of spatial gradient, requires newer degrees of freedom for each increasing order.
\end{abstract}
\maketitle

\section{Introduction}
The widespread applicability and reasonable success of fluid dynamics were never able to fully condone the pathologies arising in its theoretical formulation. Being an effective theory, which for a nonequilibrium system has been developed in terms of subleading derivative corrections, the order truncation (of derivatives) has always come with issues.
The relativistic fluid equations truncated at the first subleading order in the derivative expansion (the relativistic Navier-Stokes (N-S) equation \cite{Landau,Eckart:1940te}), admit solutions with superluminal signal propagation violating causality \cite{Hiscock:1983zz,Hiscock:1987zz}. The well-known solution comes as the Muller-Israel-Stewart (MIS) theory \cite{Israel:1979wp,Muller:1967zza} where the dissipative fluxes like viscous tensor or diffusion flow are promoted as new fluid variables, with separate equation of motion for it.
The resulting combined set of equations predicts causal signal propagation with a finite number of derivatives truncated at a certain order \cite{Hiscock:1983zz,Olson:1990rzl,Pu:2009fj}.

However, the problem remains with the physical origin of these newly promoted `non-fluid' degrees of freedom. They are not associated with any thermodynamic conserved charges, nor do they have any equilibrium counterpart (like the fundamental fluid variables such as temperature or fluid velocity). The most disturbing fact about them lies in the fact, that the number of such newly introduced variables increases with increasing order truncation. In a recent analysis \cite{Brito:2021iqr} it has been shown that for a third-order theory, even shear viscous tensor does not suffice to be treated as a new degree of freedom and leads to acausality. Restoration of causality requires more nonequilibrium variables to be treated as the new degrees of freedom. These facts indicate that the introduction
of new degrees of freedom in a higher-order theory and its connection with causality need in-depth investigation. In \cite{Mitra:2023ipl} it has been argued that the MIS theory includes all orders of derivative corrections and the all-order resummation of these gradient contributions is equivalent to introducing new ‘non-fluid’ degrees of freedom. But one still needs a proper derivation of a higher order (out-of-equilibrium) hydrodynamic theory from a first principle microscopic description and track down the source of these new degrees of freedom. The underlying microscopic theories are always known to be free from pathologies concerning subluminal signal propagation. Then the process of coarse-graining must have to do something with the rising causality related issues and the introduction of `non-fluid' variables as new degrees of freedom to deal with them. In this work, the exact (no truncation considered) dissipative, relativistic hydrodynamics has been derived from the covariant kinetic theory. Starting from the Boltzmann transport equation, the full expression of dissipative fluxes has been derived in a conformal system with no external charges. The resulting dissipative correction includes all order of derivative contributions as expected. But it is observed that the preservation of causality never allows any temporal order truncation and the infinite sums over the temporal derivatives are actually folded as the new degrees of freedom. Order truncation in hydrodynamic theory only makes sense in terms of spatial gradients. It is further observed that with increasing spatial derivative order, the power over the associated infinite temporal sum also increases which is essential for causality. This is the reason for requiring newer degrees of freedom with each increasing spatial gradient order in a hydrodynamic theory. Throughout the analysis, the calculations are kept up to linear in derivative operations considering near equilibrium scenario (which means any nonlinear term that contains product of two or more derivatives over field variables, such as $(\partial_{\mu}u^{\alpha})(\partial_{\nu}u^{\beta})$ have been neglected, whereas linear terms like $\partial_{\mu_{1}}\partial_{\mu_{2}}\cdots\partial_{\mu_{n}}u^{\nu}$ are always included in the calculations). Since the relevant derivative corrections in this analysis (in the context of causality) are always linear (however high $n$ is), this linearity treatment does not impact the generality of the outcome.

\section{Notations and convention}
Throughout the manuscript, natural unit ($\hbar = c = k_{B} = 1 $) and flat space-time with mostly negative metric signature $g^{\mu\nu} = \text{diag}\left(1,-1,-1,-1\right)$ is being used.
$\varepsilon, T, P$, $u^{\mu}$ and $\pi^{\mu\nu}$ are the energy density, temperature, pressure, hydrodynamic four-velocity and shear viscous flux respectively. The calculations are done in the local rest frame that is defined as $u^{\mu}=(1,0,0,0)$. The space projection operator is defined as $\Delta^{\mu\nu}=g^{\mu\nu}-u^{\mu}u^{\nu}$, which is orthogonal to $u^{\mu}$. $\Delta^{\mu\nu\alpha\beta}=\frac{1}{2}\Delta^{\mu\alpha}\Delta^{\nu\beta}+\frac{1}{2}\Delta^{\mu\beta}\Delta^{\nu\alpha}
-\frac{1}{3}\Delta^{\mu\nu}\Delta^{\alpha\beta}$ is denoted as the traceless projection operator orthogonal both to $u_{\mu}$ and $\Delta_{\mu\nu}$. Any space-projected vector is defined as $A^{\langle\mu\rangle}=\Delta^{\mu\nu}A_{\nu}$ and any rank-2, traceless, symmetric tensor is defined as, $A^{\langle\mu}B^{\nu\rangle}=\Delta^{\mu\nu}_{\alpha\beta}A^{\alpha}B^{\beta}$. The used derivative operators include, $D=u^{\mu}\partial_{\mu}$ as the covariant time derivative, $\nabla^{\mu}=\Delta^{\mu\nu}\partial_{\nu}$ as the spatial gradient and
$\sigma^{\mu\nu}=\nabla^{\langle\mu}u^{\nu\rangle}$ as the traceless, symmetric velocity gradient with $\partial_{\mu}$ as the 4-space-time derivative. The single particle distribution function $f(x,p)$ is defined in terms of particle momenta $p^{\mu}$ and space-time coordinate $x^{\mu}$ with $f_0 (x,p)$ as the equilibrium distribution and $\delta f(=f-f_0)$ being the out of equilibrium correction. To express $f_0$, we adopted Boltzmann distribution as $f_0(x,p)=exp[-p^{\mu}u_{\mu}/T]$ in absence of any conserved charges. $E_p(=p^{\mu}u_{\mu})$ is the single particle energy. The scaling notation $\tilde{a}$ denotes $a/(\varepsilon+P)$. The phase-space factor for moment integral is defined as $d\Gamma_p=\frac{d^3p}{(2\pi)^3 p^0}$ and $dF_p=d\Gamma_p f_0$. $\eta$ is defined as the coefficient of shear viscosity.

\section{Exact expression of viscous flux from kinetic theory}
Relating the viscous effects with the distortion of the constituent particle's momentum distribution, the idea here is to express the viscous fluxes as the moments of the out-of-equilibrium single particle distribution function. For that, we need an equation of motion that solves the out-of-equilibrium distribution function as a function of space-time and particle momenta.
In this analysis, we take the recourse of the relativistic transport equation which describes the space-time evolution of the single particle momentum distribution function in terms of a microscopic collision term as \cite{Degroot},
\begin{align}
\tilde{p^{\mu}}\partial_{\mu}f=-\frac{\tilde{E}_p}{\tau_R}\delta f=-\frac{\tilde{E}_p}{\tau_R}\left(f-f_0\right)~,
\label{RTE-1}
\end{align}
where the collision kernel has been linearized in terms of the well known relaxation time approximation (RTA) \cite{RTA}. $\tau_R$ is the relaxation time scale for $f$ to restore its equilibrium value $f_0$ via collisions.
Eq.\eqref{RTE-1} solves the out-of-equilibrium distribution function as,
\begin{align}
 \delta f=-\frac{\left[\frac{\tau_R}{\tilde{E}_p}\tilde{p}^{\mu}\partial_{\mu}\right]f_0}
 {\left[1+\frac{\tau_R}{\tilde{E}_p}\tilde{p}^{\mu}\partial_{\mu}\right]}~.
 \label{deltaf-1}
\end{align}
Only allowing the linearized derivative operations as mentioned earlier (any products of derivatives are neglected considering near equilibrium scenario) and in the absence of any conserved charges, the out-of-equilibrium distribution function takes the form,
\begin{align}
 \delta f=\frac{\frac{\tau_R}{\tilde{E}_p}f_0\left[\tilde{p}^{\langle\mu}\tilde{p}^{\nu\rangle}\sigma_{\mu\nu}-\frac{1}{(\varepsilon+P)}\tilde{E}_p\tilde{p}_{\langle\nu\rangle
 }\nabla_{\rho}\pi^{\nu\rho}\right]}{\left[1+\frac{\tau_R}{\tilde{E}_p}\tilde{p}^{\lambda}\partial_{\lambda}\right]}~.
 \label{deltaf-2}
\end{align}
Plugging Eq.\eqref{deltaf-2} in the moment equation of $\pi^{\mu\nu}$ as follows,
\begin{align}
 \frac{\pi^{\alpha\beta}}{T^2}&=\int d\Gamma_p \tilde{p}^{\langle\mu} \tilde{p}^{\nu\rangle}\delta f~,
 \label{pi1}
\end{align}
we got the following equation for the shear viscous flux,
\begin{align}
 \frac{\pi^{\alpha\beta}}{T^2}&=
 \int dF_p\frac{\frac{\tau_R}{\tilde{E}_p}\tilde{p}^{
 \langle\alpha}\tilde{p}^{\beta\rangle} \tilde{p}^{
 \langle\mu}\tilde{p}^{\nu\rangle}\sigma_{\mu\nu}}{\left[1+\tau_R D+\frac{\tau_R}{\tilde{E}_p}\tilde{p}^{\langle\lambda\rangle}\nabla_{\lambda}\right]}\nn
&-\frac{\tau_R}{(\varepsilon+P)}
 \int dF_p\frac{\tilde{p}^{
 \langle\alpha}\tilde{p}^{\beta\rangle} \tilde{p}_{\langle\mu\rangle}\nabla_{\nu}\pi^{\mu\nu}}{\left[1+\tau_R D+\frac{\tau_R}{\tilde{E}_p}\tilde{p}^{\langle\lambda\rangle}\nabla_{\lambda}\right]}~.
 \label{pi2}
\end{align}
Eq.\eqref{pi2} is the exact expression of $\pi^{\mu\nu}$, no order truncation has been made in the definition \eqref{pi1}.
Here we make an important observation. The term in the derivative operator appearing in the denominator of Eq.\eqref{pi2} can be approximated as,
\begin{align}
 \tau_R D+\frac{\tau_R}{\tilde{E}_p}\tilde{p}^{\langle\lambda\rangle}\nabla_{\lambda}~~\sim~~\frac{\lambda_{\text{mfp}}}{L}=K_n~.
\end{align}
$\lambda_{\text{mfp}}$ is mean free path (microscopic length scale) of the system and $L$ is the macroscopic length scale over which the
thermodynamic state variables considerably change. The ratio $K_n$ is termed as the Knudsen number \cite{Degroot,Cercignani} that decides the region of validity of the resulting coarse-grained (hydrodynamic) theory. $K_n<1$ is the known limit for hydro to be valid where it is defined in terms of systematic build up of gradients of the fluid variables. Within this limit, i.e. $|\tau_R D+\frac{\tau_R}{\tilde{E}_p}\tilde{p}^{\langle\lambda\rangle}\nabla_{\lambda}|<1$, the denominator of Eq.\eqref{pi2} can be expanded as an infinite derivative sum series operating on thermodynamic force terms ($\sigma^{\mu\nu}$ and $\nabla_{\nu}\pi^{\mu\nu}$). A little bit of cumbersome but straightforward mathematical analysis can appropriately sum these derivatives to obtain the following expression of $\pi^{\mu\nu}$,
\begin{align}
 \frac{\pi^{\alpha\beta}}{T^2}&=
 \sum_{m=0}^{\infty}\left\{\frac{\tau_R}{1+\tau_R D}\right\}^{2m+1}\nn
 \times&\left[\int \frac{dF_p}{\left(\tilde{E}_p\right)^{2m+1}}\tilde{p}^{\langle\alpha}\tilde{p}^{\beta\rangle}
 \tilde{p}^{\langle\mu}\tilde{p}^{\nu\rangle}\tilde{p}^{\langle\lambda_1\rangle}\cdots\tilde{p}^{\langle\lambda_{2m}\rangle}\right]\nabla_{\lambda_1}\cdots\nabla_{\lambda_{2m}}\sigma_{\mu\nu}\nn
 +&\frac{1}{(\varepsilon +P)}\sum_{n=0}^{\infty}\left\{\frac{\tau_R}{1+\tau_R D}\right\}^{2n+2}\nn
 \times&\left[\int \frac{dF_p}{\left(\tilde{E}_p\right)^{2n+1}}\tilde{p}^{\langle\alpha}\tilde{p}^{\beta\rangle}\tilde{p}_{\langle\mu\rangle}\tilde{p}^{\langle\lambda_1\rangle}\cdots\tilde{p}^{\langle\lambda_{2n+1}\rangle}\right]\nabla_{\lambda_1}\cdots\nabla_{\lambda_{2n+1}}\nabla_{\nu}\pi^{\mu\nu}.
 \label{pi3}
\end{align}
This is the exact expression of $\pi^{\alpha\beta}$ that includes all orders of spatial as well as temporal derivatives.
The infinite sum over the temporal derivatives forms a closed structure that creates relaxation operator like forms ($1+\tau_R D$) in the denominator of \eqref{pi3}. However, the resummation of the spatial derivatives is not trivial because of the prefactor contribution in each term resulting from the moment integrals. The crucial observation here is that, in each term of Eq.\eqref{pi3}, the order of spatial gradient in the numerator exactly agrees with the order of temporal derivative in the denominator. We will return to the significance of the observation once the final expression of $\pi^{\mu\nu}$ is obtained. In order to do that, all is left to perform the moment integrals in Eq.\eqref{pi3}. For a massless system, the moment integrals are given by the following formula \cite{Degroot},
\begin{align}
\Gamma^{\nu_1\cdots\nu_n}&=\int dF_p \tilde{p}^{\nu_1}\tilde{p}^{\nu_2}\cdots\tilde{p}^{\nu_n}=\sum_{l=0}^{[n/2]}a_{nl}\left(\Delta u\right)_{nl}~,\nn
a_{nl}&=(-1)^l \frac{T^2}{2\pi^2}\frac{n!(n+1)!}{(n-2l)!(2l+1)!}~,\nn
\left(\Delta u\right)_{nl}&=\frac{1}{n!}\sum_{\cal{P}}\Delta^{\nu_1\nu_2}\cdots\Delta^{\nu_{2l-1}\nu_{2l}}u^{\nu_{2l+1}}\cdots u^{\nu_{n}}~,
\label{moment}
\end{align}
where the summation is extended over all permutations ${\cal{P}}$ of the indices $\nu_n$. Using the prescription given in \eqref{moment}, estimating the moments of Eq.\eqref{pi3} leads to some nasty and cumbersome algebra, but it is still analytically doable. Pursuing that, the final result of the exact (no order truncation done) shear viscous tensor is given as the following,
\begin{align}
\frac{\pi^{\alpha\beta}}{4!P}=&
\sum_{m=0}^{\infty}(-1)^m\frac{1}{(2m+5)}\frac{1}{(2m+3)}\frac{1}{(2m+1)}\left\{\frac{\tau_R}{1+\tau_R D}\right\}^{2m+1}\nn
&\times\bigg[\left(\nabla^2\right)^m\sigma^{\alpha\beta}+(4m)\left(\nabla^2\right)^{m-1}\nabla^{\langle\alpha}\nabla^{\nu}\sigma^{\beta\rangle}_{\nu} \nn
&+(2m)(m-1)\left(\nabla^2\right)^{m-2}\nabla^{\langle\alpha}\nabla^{\beta\rangle}
\nabla_{\langle\mu}\nabla_{\nu\rangle}\sigma^{\mu\nu}
\bigg]\nn
&+\frac{1}{\left(\varepsilon+P\right)}
\sum_{n=0}^{\infty}(-1)^n\frac{1}{(2n+5)}\frac{1}{(2n+3)}\left\{\frac{\tau_R}{1+\tau_R D}\right\}^{2n+2}\nn
&\times\left[\left(\nabla^2\right)^n\nabla^{\langle\alpha}\nabla^{\nu}\pi^{\beta\rangle}_{\nu}+n\left(\nabla^2\right)^{n-2}\nabla^{\langle\alpha}\nabla^{\beta\rangle}
\nabla_{\langle\mu}\nabla_{\nu\rangle}\pi^{\mu\nu}
\right]~.
\label{pifinal1}
\end{align}
The expression given in \eqref{pifinal1} is basically an all order differential equation satisfied by $\pi^{\alpha\beta}$, whose solution is presented by the following equation,
\begin{align}
 &\left(\frac{1+\tau_R D}{\tau_R}\right)\pi^{\mu\nu}=\nn
 &\Bigg[4!P\sum_{m=0}^{\infty}(-1)^m\frac{1}{(2m+5)}\frac{1}{(2m+3)}\frac{1}{(2m+1)}\left(\frac{\tau_R}{1+\tau_R D}\right)^{2m}\nn
 &~~~~~~~~~~~~\times\Bigg\{\left(\delta^{\alpha}_{\mu}\delta^{\beta}_{\nu}\right)\nabla^{2m}+(4m)\nabla^{2m-2}\Delta^{\alpha\beta\lambda}_{\mu}\nabla_{\nu}\nabla_{\lambda}\nn
 &~~~~~~~~~~~~+(2m)(m-1)\nabla^{2m-4}\nabla^{\langle\alpha}\nabla^{\beta\rangle}
 \nabla_{\langle\mu}\nabla_{\nu\rangle}\Bigg\}\sigma^{\mu\nu}\Bigg]~~{\big{/}}\nn
&\Bigg[\delta^{\alpha}_{\nu}\delta^{\beta}_{\nu}-6  \sum_{n=0}^{\infty}(-1)^n\frac{1}{(2n+5)}\frac{1}{(2n+3)}\left(\frac{\tau_R}{1+\tau_R D}\right)^{2n+2}\nn
 &~~~~~~~~~~~~\times\Bigg\{\nabla^{2n}\Delta^{\alpha\beta\lambda}_{\mu}\nabla_{\nu}\nabla_{\lambda}
+n\nabla^{2n-2}\nabla^{\langle\alpha}\nabla^{\beta\rangle}
 \nabla_{\langle\mu}\nabla_{\nu\rangle}\Bigg\}\Bigg]~.
 \label{pifinal2}
\end{align}
Eq.\eqref{pifinal1} and \eqref{pifinal2} are the main results of this work. They include both temporal and spatial derivatives up to infinite order to depict the nonequilibrium effects. The prime observations from these results are that, (i) the resummed temporal derivatives form nonlocal operators (time derivative appearing in the denominator) which further operate on the thermodynamic forces (spatial derivatives operating on field variables like velocity here), and (ii) with increasing order of spatial derivatives of the thermodynamic force terms, the power of the nonlocal operator also systematically increases. Since for a conformal, chargeless system, the only dissipative quantity is the shear viscous tensor, the order of spatial derivatives comes in odd powers only.
Now such an all-order theory needs to be truncated for practical purposes like phenomenological applications. Next, we will see what order truncation means in an out-of-equilibrium theory and what connection it has with the causality of the theory.

\section{Derivative truncation in a relativistic dissipative theory}
The conservation law for an out-of-equilibrium energy- momentum tensor $T^{\mu\nu}$ (in a conformal system with no conserved charges) is given by,
\be
\partial_{\mu}T^{\mu\nu}=0~,~~~~~~T^{\mu\nu}=\varepsilon u^{\mu}u^{\nu}-P\Delta^{\mu\nu}+\pi^{\mu\nu}~.
\label{enmom}
\ee
Linearizing \eqref{enmom} for small perturbations of fluid variables around the hydrostatic equilibrium in the local rest frame ($\psi(x,t)=\psi_0+\delta\psi(x,t)$) and then expressing them in plane wave solutions $(\delta\psi(t,x)\rightarrow e^{i(\omega t-kx)} \delta\psi(\omega,k)$, with $k^{\mu}=(\omega,\k,0,0))$, one obtains the dispersion relations $F(\omega,\k)=0$.
Upon solving this, the frequency modes become available for further analysis such as causality. Recently, in \cite{Hoult:2023clg}, a certain property of $F(\omega,\k)$ concerning the number of modes has been mentioned as a required condition for causality. It states that,
\be
{\cal{O}}_{\omega}[F(\omega,\k\neq0)]= {\cal{O}}_{\vert\k\vert}[F(\omega=a\vert\k\vert,\k=\bf{b}\vert{\k}\vert]~,
\label{caus}
\ee
with $a$ as a non-zero, real scalar constant, $\bf{b}$ as a real unit vector and ${\cal{O}}_x$ denoting the order of the polynomial in the variable $x$. Further details can be found in \cite{Roy:2023apk}. We will now discuss how the condition \eqref{caus} becomes relevant in the context of the results obtained in \eqref{pifinal1} and \eqref{pifinal2} and in order truncation (of gradients) for a causal dissipative theory as a whole.

In terms of the field perturbations, for the shear channel (identical analysis can be trivially done for sound channel as well) Eq.\eqref{enmom} becomes,
\be
(\varepsilon+P)~\omega ~\delta u^{y}-\k~\delta\pi^{xy}=0~.
\label{sheardisp}
\ee
Now replacing $\pi^{\mu\nu}$ from \eqref{pifinal1} and \eqref{pifinal2} ($\delta\pi^{\mu\nu}=\pi^{\mu\nu}$, since dissipative quantities have no global equilibrium counterpart) in \eqref{sheardisp}, we will obtain $F(\omega,\k)=0$ that must satisfy \eqref{caus} to respect causality of the theory. We see that in each term of \eqref{pifinal1} and \eqref{pifinal2}, the order of spatial gradients in the numerator (in the force terms) and the order of temporal derivatives in the denominator (in the nonlocal terms) exactly agrees with each other. As a result, after replacing $\pi^{\mu\nu}$ in \eqref{sheardisp}, any truncation of $n$ and $m$ values in \eqref{pifinal1} and \eqref{pifinal2} perfectly satisfies causality condition \eqref{caus}. Here we state the prime argument of the current finding. The order truncation in any dissipative, relativistic hydrodynamic theory must be purely based on the truncation of spatial gradients. The temporal gradients must include all orders (up to infinity)
for any given spatial order. It can be clearly seen that in the expression of $\pi^{\mu\nu}$ if the nonlocal operator $(1+\tau_{R}D)$ is expanded in terms of temporal derivatives operating on thermodynamic forces, and that expansion is truncated at any finite order (however high it is), the causality condition \eqref{caus} is surely violated. So any causal, dissipative theory is basically an all-order theory (over temporal derivatives), and higher order truncation only makes sense in terms of certain spatial gradient order. Keeping up to first order in spatial gradients ($m=0$, no $n$ contribution) we obtain the convention MIS theory
$(1+\tau_R D)\pi^{\alpha\beta}=2\eta^{\alpha\beta}$ \cite{Israel:1979wp,Muller:1967zza}.
Keeping up to the third order of spatial gradients ($m=0,1$ and $n=0$), one can obtain the results of \cite{Jaiswal:2013vta} at the linearized level. Although the abrupt truncation of temporal derivatives as well in \cite{Jaiswal:2013vta} creates the causality issues which have been addressed in \cite{Brito:2021iqr} and remedied by introducing another new degree of freedom apart from $\pi^{\mu\nu}$. This fact of requiring more and more degrees of freedom for higher-order theories has been addressed in the next section.

\section{New degrees of freedom in a higher order theory}
So we see that the relativistic, dissipative hydrodynamics (Eq.\eqref{pifinal1} and \eqref{pifinal2}) is a nonlocal theory in time because of the temporal operator $(1+\tau_R D)$ appears in the denominator of each term in $\pi^{\mu\nu}$. Such nonlocalities are not welcome for practical purposes, especially since numerical simulations can generate pathologies with such terms. Generically, a nonlocal set of equations can be recast into a local set of equations by `integrating in' new `non-fluid' variables. In other words, the nonlocalities in \eqref{pifinal1} and \eqref{pifinal2} can be absorbed by introducing new degrees of freedom apart from the fundamental fluid degrees of freedom (temperature and fluid velocity). Each such new degrees freedom must have their own differential equations to be solved. Following that prescription, Eq.\eqref{pifinal1} and \eqref{pifinal2} can be recast as a local set of equations with an infinite number
of degrees of freedom in the following fashion,
\begin{align}
 \left(1+\tau_R D\right)\pi^{\alpha\beta}&=2\eta\sigma^{\alpha\beta} +\rho_1^{\alpha\beta}~,\nn
\left(1+\tau_R D\right)^2\rho_{1}^{\alpha\beta}&=-\eta\tau_R^2\left[\frac{2}{7}\nabla^2\sigma^{\alpha\beta}+\frac{12}{35}\nabla^{\langle\alpha}\nabla^{\nu}\sigma^{\beta\rangle}_{\nu}\right]+\rho_2^{
\alpha\beta}~,\nn
\left(1+\tau_R D\right)^2\rho_{2}^{\alpha\beta}&=\eta\tau_R^4\Bigg[\frac{2}{21}\nabla^4\sigma^{\alpha\beta}+\frac{32}{105}\nabla^2\nabla^{\langle\alpha}\nabla^{\nu}\sigma^{\beta\rangle}_{\nu}\nn
&~~~~~~~~~+\frac{4}{105}\nabla^{\langle\alpha}\nabla^{\beta\rangle}\nabla_{\langle\mu}\nabla_{\nu\rangle}\sigma^{\mu\nu}\nn
&~~~~~~~~~-\frac{16}{35}\Delta^{\alpha\beta}_{ab}\Delta^{b\nu}_{cd}\nabla^a\nabla^c\nabla_{\nu}\nabla_{\rho}\sigma^{\rho d}\Bigg]+\rho_3^{\alpha\beta~,}\nn
\vdots
\label{dof}
\end{align}
Here, under RTA, $\eta=(4/5)P\tau_R$ has been used \cite{Mitra:2020gdk}. From \eqref{dof} it can be seen that for each higher order of spatial gradient truncation, a new degree of freedom needs to be introduced. Otherwise, the causality of the theory is bound to be compromised. That is why, for a theory having only up to first order of spatial gradient (force term only contains $\sigma^{\mu\nu}$) only one degree of freedom $\pi^{\mu\nu}$ is sufficient. If the order of spatial gradient increases to three \cite{Jaiswal:2013vta}, another new degree of freedom needs to be introduced in order to maintain causality \cite{Brito:2021iqr} and so on.

This infinite loop of newer degrees of freedom with each increasing spatial gradient order of the same theory is certainly unsettling. The reason may be traced in the following way. As very nicely remarked in \cite{Bemfica:2017wps}, the underlying microscopic descriptions of a hydrodynamic theory such as kinetic theory based on the Boltzmann equation are always pathology free. These causality related issues are basically the limitations or artifacts of a particular coarse-graining method. So the information of the causality preservation of the microscopic theory is encapsulated in the hydrodynamic derivation in the form of the nonlocality over time. This in order to make the theory local is absorbed via newer degrees of freedom. So in light of that argument, these newer degrees of freedom are exactly what they appear. These quantities are neither associated with any thermodynamic conserved charges nor have any equilibrium counterparts (like the fundamental fluid degrees of freedom such as $T$ and $u^{\mu}$). They are just the mathematical artifacts essential for a causal, local theory to be pathology free at higher orders.

A remark is needed to be made here. As mentioned in \cite{Bhattacharyya:2024tfj,Bhattacharyya:2023srn}, this method of introducing new non-fluid degrees of freedom is highly non-unique. Several parallel techniques can be opted for it. For example, the way newer degrees of freedom are introduced in \cite{Brito:2021iqr} is a bit different than the method has been depicted here in \eqref{dof}.

\section{Conclusion}
In this work, the emergence of new degrees of freedom in a (finite) higher-order relativistic, viscous hydrodynamic theory has been probed. Derived from a microscopic theory (covariant kinetic theory in the present analysis), it has been shown that the information of causality is retained in the infinite sum of the temporal derivatives which appears as the nonlocal operators in the resulting hydrodynamic theory. In order to make it a local theory, these nonlocalities are recast into a local set of equations by introducing new ‘non-fluid’
variables. These ‘non-fluid’ variables are treated as the new degrees of freedom for the theory. With higher order of spatial gradients, the power over these nonlocal operators simultaneously increases in order to preserve causality. Hence, for a theory truncated at a higher order of spatial gradient, more and more new degrees of freedom are required accordingly.

This issue of degrees of freedom has been recently taken care in a first-order, causal-stable theory (popularly known as the Bemfica-Disconzi-Noronha-Kovtun (BDNK) theory) \cite{Bemfica:2017wps,Kovtun:2019hdm} where the fluid description suffices in terms of primary fluid variables ($T,u^{\mu}$) only. The price is paid in the definition of fluid variables since they are defined by arbitrary out-of-equilibrium contributions. In \cite{Bhattacharyya:2024tfj} it has been argued that we have to make a consensus between the two. However, introducing new degrees of freedom does not have any issue with hydrodynamic numerical simulations (apart from the additional differential equations to solve) and MIS has been largely successful in phenomenological explanation of the heavy-ion data from high energy experiments \cite{Romatschke:2017ejr,Huovinen:2006jp}. This work has been an effort to investigate the physical (mathematical) meaning of these new degrees of freedom appearing in a higher-order theory.

\section{Acknowledgements}
I duly acknowledge Sayantani Bhattacharyya for conceptual discussions and valuable inputs.
For the ﬁnancial support S.M. acknowledges the Department of Atomic Energy, India.

\end{document}